\documentclass[twocolumn,showpacs,prc,floats,amsmath,amssymb]{revtex4}
\usepackage{graphicx}
\usepackage{dcolumn}
\newcommand{\be}{\begin{equation}}
\newcommand{\ee}{\end{equation}}
\newcommand{\ba}{\begin{eqnarray}}
\newcommand{\ea}{\end{eqnarray}}
\newcommand{\sba}{\begin{subequations}}
\newcommand{\sea}{\end{subequations}}
\newcommand{\barr}{\begin{array}}
\newcommand{\earr}{\end{array}}
\newcommand{\nn}{\nonumber \\}
\newcommand{\bm}{\begin{mathletters}} 
\newcommand{\eml}{\end{mathletters}}

\newcommand{\dl}{\delta}

\def\C {{{\cal C}}}
\def\D {{{\cal D}}}

\def\G {{{\cal G}}}

\def\H {{{\cal H}}}

\def\M {{{\cal M}}}
\def\N {{{\cal N}}}

\def\X {{{\cal X}}}
\def\Y {{{\cal Y}}}

\begin{document}
 
\title{Coupled Self-Consistent RPA Equations for Even and Odd Particle Numbers. \\
Tests with Solvable Models.
}
\date{\today}

\author{M. Jema\"i}
\affiliation{Laboratoire des mat\'eriaux avanc\'es et ph\'enom\`enes quantiques, 
FST, Universit\'e Tunis El-Manar 2092 El-Manar, Tunis, Tunisia.}
\affiliation{ISSATM, Universit\'e de Charthage, Avenue de la R\'epublique P.O. Box 77 - 1054 Amilcar, Tunis, Tunisia.}
\email{mohcen.jemai@issatm.u-carthage.tn}
\author{P. Schuck}
\affiliation{  Institut de Physique Nucl\'eaire d'Orsay, Universit\'e Paris-Sud, CNRS--IN2P3 \\
               15, Rue Georges Clemenceau, 91406 Orsay Cedex, France.}
\affiliation{ Univ. Grenoble Alpes, CNRS, LPMMC, 38000 Grenoble, France }
\email{schuck@ipno.in2p3.fr}

\begin{abstract}
Coupled equations for even and odd particle number correlation functions are set up via the equation of motion method. For the even particle number case this leads to self-consistent RPA (SCRPA) equations already known from the literature. From the equations of the odd particle number case the single particle occupation probabilities are obtained in a self-consistent way. This is the essential new procedure of this work. Both, even and odd particle number cases are based on the same correlated vacuum and, thus, are coupled equations. Applications to the Lipkin model and the 1D Hubbard model give very good results.
\end{abstract}
\maketitle

\section{Introduction}

Developments of Many-Body approaches for strongly correlated systems is an active field of research. In the past we developed an RPA theory which goes beyond the standard one and which is based on a correlated ground state. To lowest order this leads to RPA equations where the single particle (s.p.) occupation numbers are not the uncorrelated (Hartre-Fock (HF)) ones but correlated ones which are obtained in a self-consistent way from the RPA solution. It is known as 're-normalized' RPA (r-RPA) \cite{DSD05,SchaferP}. 
However, in general the RPA equations contain additionally vertex (e.g., screening) corrections
which also can be obtained self-consistently from the RPA solution. The whole procedure has been dubbed Self-Consistent RPA (SCRPA) \cite{jemai13} and references in there. SCRPA can also be seen as a sub-product of an even more general approach which is the Time-Dependent Density Matrix (TDDM) theory based on a decoupling of the BBGKY hierarchy of one, two, etc.  correlation functions \cite{SchuToh, tddm-scrpa}. In the past, there was always a certain debate how to include the single particle (s.p.) occupations into the SCRPA scheme. Since the latter are s.p. quantities and the RPA gives rise to two body correlations, it was not completely evident in which way to close the system of equations. However, already Rowe promoted the so-called particle number operator method to obtain self-consistent s.p. occupations \cite{Row68}. Later this was further elaborated by F. Catara \cite{Catara} and it has become known as the 'Catara method' since then. This method mostly works quite well \cite{jemai05} but also fails more or less in particular cases \cite{jemai13}. It is for this reason that in this work we elaborate a different scheme which seems to be more natural since it is also based on the equation of motion (eom) method and employs the so-called odd particle number RPA (o-RPA) recently proposed by one of the authors plus collaborators \cite{TohSchu13}. The latter can also be reformulated as a Dyson equation for the s.p. Green's function (GF) with a self-energy obtained from the eom for the $2$particle-$1$hole (2p-1h) and 2h-1p GF. Since both, the SCRPA and o-RPA will be based on the same correlated vacuum, naturally even and odd particle number channels become coupled. We may coin this scheme eo-SCRPA. We will apply this approach to two exactly solvable model cases: the Lipkin model and the 1D Hubbard model. In both cases the results turn out to be promising.

The paper is organized as follows: in section II we present the general theory. In section III, applications to the Lipkin and Hubbard models are given. Section IV contains the conclusions and details of our procedures are presented in the Appendices.

\section{General Theory}
\label{GeneTheo}
As mentioned in the introduction, we will base our approach on the coupling of the even and odd particle number eom. The latter will be obtained from the following ansatz 
\begin{eqnarray}
q^\dag_{\mu} =\sum_h x^\mu_{h} a_{h} +\sum_{pp'h'} U^\mu_{pp'h'} a^\dagger_{h'} a_{p'} a_{p}~,
\nn
q^\dag_{\rho}=\sum_p x^\rho_p a^\dagger_p +\sum_{p'h'h} U^\rho_{p'h'h} a^\dagger_{h}a^\dagger_{h'}a_{p'}~.
\label{odd-exc-Op}
\end{eqnarray}
where the indices 'p, h' refer to s.p. states 'above' and 'below' the Fermi surface, respectively, and $a^\dagger_k, a_k$ are the fermion creation and annihilation operators. The equation for the s.p. basis in which the equations will be worked out will be given below. Our 'quasi-particle' operator in (\ref{odd-exc-Op}) has the good quality that its destructor exactly kills the so-called Coupled-Cluster-Doubles (CCD) wave function:
\be q_{\mu}|Z\rangle = 0, \ee
with
\ba
|Z\rangle &=& \exp\left(\sum_{pp'hh'}\frac{1}{4}Z_{pp'hh'} a^\dagger_{p} a^\dagger_{p'} a_{h'}a_h\right)|\rm HF\rangle ,\label{CCD}
\ea
where $|\rm HF\rangle$  is the Hartree-Fock (HF) Slater determinant and the amplitudes
must full-fill the following relations 
\ba
\sum_h x^{\mu *}_{h} Z_{pp'hh'} &=& U^{\mu *}_{pp'h'} ,~~ \sum_p x^{\rho *}_p Z_{pp'hh'} 
=U^{\rho *}_{p'h'h}.~
\ea
The coefficients will be determined from the minimisation of a sum rule for the average s.p. energy 
\ba
\lambda_\mu &=& \frac{1}{2} \frac{\langle\{q_\mu,[H,q^\dag_\mu]\}\rangle}{\langle\{q_\mu,q^\dag_\mu\}\rangle}
\nn
&=&\frac{1}{\langle 0|\{q_{\mu},q_{\mu}^\dag\}|0\rangle}
\sum_{\alpha} (E^{N+1}_{\alpha}-E^{N}_0)|\langle 0|q_{\mu}|\alpha\rangle |^2 
\label{minim-odd}
\ea
and equivalently for $q^\dag_{\rho}$. The even particle number equation relies on the usual RPA excitation operator \cite{RingSchuck}
\begin{equation}
Q^\dag_\nu =\sum_{ph} X^\nu_{ph} a^\dagger_{p}a_{h} -Y^\nu_{ph} a^\dagger_{h}a_{p}~.
\label{evenexctOpRPA}
\end{equation}
The $X,~Y$ amplitudes are the solutions of another sum-rule defining an average excitation energy of the even systems
\ba
\Omega_{\nu} &=& \frac{1}{2}\frac{\langle 0|[Q_{\nu},[H,Q_{\nu}^\dag]]|0\rangle}
{\langle 0|[Q_{\nu},Q_{\nu}^\dag]|0\rangle}
\nn
&=&\frac{1}{\langle 0|[Q_{\nu},Q_{\nu}^+]|0\rangle}
\sum_{\mu} (E_{\mu}-E_0)|\langle 0|Q_{\nu}|\mu\rangle |^2 ~.
  \label{minim-even}
\ea
The destruction operator $Q_{\nu}$ does not exactly kill the CCD ground state without introducing a generalization \cite{jemai13} but it kills it to very good approximation as studies of model cases have shown \cite{jemai13}. Therefore the even particle number equation (SCRPA) is the only point where the approach is not entirely consistent though the theory remains very performent as we will see below in the Application section. We, thus, will henceforth always suppose that
\ba 
Q|Z\rangle = 0~. 
\ea
The SCRPA equation corresponding to the minimisation of the mean excitation energy $\Omega_{\nu}$ in (\ref{minim-even}) can be written as
\ba
\left(\begin{array}{cc} A &  B \\ -B^* & -A^* \end{array} \right) 
\left( \begin{array}{c} X^{\nu } \\ Y^{\nu }\end{array} \right) 
=\Omega _{\nu} \left( \begin{array}{c} X^{\nu } \\ Y^{\nu } \end{array} \right) 
\label{RPAeqGen}
\ea
\noindent
with the normalisation of the amplitude given as usual by

\ba
\sum_{ph}\left(\vert X^{\nu}_{ph}\vert ^2 - \vert Y^{\nu}_{ph}\vert ^2\right) = 1\label{normlisXY}
\ea
and
\ba
A_{ph,p'h'} &=&  \frac{\langle [a^{\dag}_ha_p,[H,a^{\dag}_{p'}a_{h'}]] \rangle }{ \sqrt{n_h-n_p}\sqrt{n_{h'}-n_{p'}}} ~,
\nonumber \\ 
B_{ph,p'h'} &=& -\frac{\langle [a^{\dag}_ha_p,[H,a^{\dag}_{h'}a_{p'}]] \rangle }{ \sqrt{n_h-n_p}\sqrt{n_{h'}-n_{p'}}}
\label{elemt-mat}
\ea
\noindent
where $\langle ...\rangle = \langle Z|...|Z\rangle/\langle Z|Z\rangle$.
The SCRPA equations are well documented in the literature \cite{DSD05,jemai05,tddm-scrpa,jemai13} and we will not repeat their explicit form here. Let us simply say that $A$ and $B$ are functional of one and two particle density matrices   when the Hamiltonian of the system is given by 
\begin{equation}
  H \,=\, \sum \limits_{kk'} t_{kk'} \, a^{\dagger}_{k} a_{k'}
  \;+\; \frac{1}{4} \sum \limits_{klmn} \bar{v}_{klmn} \,
  a^{\dagger}_{k} a^{\dagger}_{l} a_{n} a_{m}~.
  \label{HFermion}
\end{equation}
The first part of the Hamiltonian represents the kinetic energy and the second part the two body interaction with the anti-symmetrized matrix element
\[ \bar v_{klmn} = \langle kl|v|mn\rangle - \langle kl|v|nm\rangle .\]
\noindent
From the minimisation of the sum-rule in eq.(\ref{minim-odd}), we obtain two coupled equations 
\ba
&&\sum_{h'} \epsilon_{hh'} x^\mu_{h'} + \sum_{pp'h'}\C_{h,pp'h'} U^\mu_{pp'h'} =\lambda_\mu x^\mu_{h} 
\\
&&\sum_{h'}\C^{*}_{pp'h,h'} x^\mu_{h'} + \sum_{p_2p_1h_1}\D_{pp'h,p_2p_1h_1} U^\mu_{p_2p_1h_1} =\lambda_\mu U^\mu_{pp'h}\nonumber
\label{eqoddRPA}
\ea
or written as  a matrix eigenvalues equation  
\ba
\left(\begin{matrix}
\epsilon  &\C \\
\C^{\dagger} &\D
\end{matrix}\right)\left(\begin{matrix}
x^\mu\\ U^\mu
\end{matrix}\right)=\lambda_\mu \left(\begin{matrix}
x^\mu \\ U^\mu
\end{matrix}\right)
\label{odd_matrix}
\ea
with
\ba
\epsilon_{hh'} &=& \langle\{a_{h},\left[H,a^\dagger_{h'}\right]\}\rangle~~~=\epsilon_h\delta_{hh'} 
\ea
where we supposed that hitherto we work in the Mean-Field (MF) basis with diagonal s.p. MF energies $\epsilon_h, \epsilon_p$.
The matrices $\C$ and $\D$ in (\ref{eqoddRPA}) are obtained from the minimisation of the mean s.p. energy given in (\ref{minim-odd})
\ba
\C^{*}_{pp'h',h} &=&\frac{\langle \left\{a^\dagger_{h'}a_{p'}a_{p} , \left[H,a^\dagger_{h} \right]\right\}\rangle }{\sqrt{\N_{pp'h'}}}
\nn
\D_{pp'h',p_2p_1h_1} &=&\frac{\langle\left\{a^\dagger_{h'}a_{p'}a_{p}, \left[H,a^\dagger_{p_2} a^\dagger_{p_1}a_{h_1} \right]\right\}\rangle}{\sqrt{\N_{pp'h'}}\sqrt{\N_{p_2p_1h_1}}}~~~
\nn
\N_{pp'h'} &=&\langle\left\{a^\dagger_{h'}a_{p'}a_{p}, a^\dagger_{p} a^\dagger_{p'}a_{h'}\right\}\rangle
\label{elemt-matodd} 
\ea
\noindent
Equation (\ref{eqoddRPA}) is essentially already given in \cite{TohSchu13}. However, the way we solve this equation and, thus, couple it to the SCRPA of (\ref{RPAeqGen}) is novel. Let us briefly describe the procedure.
The coefficients $\C$ and $\D$ contain two and three body correlation functions. In particular they contain p-h operators  which are given by the inversion of (\ref{evenexctOpRPA}) valid because the $X, Y$ amplitudes in (\ref{RPAeqGen}) form a complete orthonormal set of states
\begin{equation}
  a^{\dag}_pa_h = \sqrt{ n_h - n_p}\sum_\nu X^\nu_{ph} Q^+_\nu + Y^\nu_{ph} Q_\nu
  \label{invQ}
\end{equation}
and its hermitian conjugate. All other correlation functions which do not contain those ph operators and which are not of the $pp'h$ type shall be discarded since they are supposedly  less important. Commuting the destructor $Q$ to the right until they hit and kill the vacuum state $|Z\rangle$ leads to expressions of diverse correlation functions which only contain s.p. occupations $n_h$ and $n_p$ and RPA amplitudes $X, Y$.
We want to call the coupled equations (\ref{RPAeqGen}-\ref{elemt-mat}) and (\ref{odd_matrix}-\ref{elemt-matodd}) 'even-odd SCRPA' (eo-SCRPA). One may find more details in the Application section below. 
This procedure to obtain the s.p. occupation numbers is the essential new point of this work. It is clear that in this way eqs. (\ref{RPAeqGen}) and (\ref{eqoddRPA}) become coupled. In our earlier publications the s.p. occupation numbers appearing in the SCRPA equations have always been obtained in a different, in our opinion less natural way. We should also say the the formal expressions of SCRPA are not altered, only the way how the s.p. occupation probabilities in there are calculated is new.

\begin{figure}
\includegraphics[width=8.5cm,height=2.5cm]{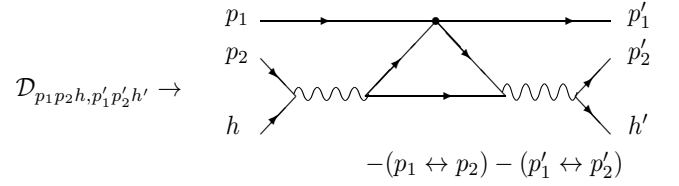}
\caption{Schematic representation of the three body interaction $\D$. The wiggly line stands for the (collective) ph-modes. The full dot represents the two body interaction. }
\label{diagramD}
\end{figure}

It may be helpful at this point to discuss for instance the matrix $\D$ in (\ref{odd_matrix},\ref{elemt-matodd}) a little more and give a graphical representation. From the double commutator in $\D$, we retain only those terms where a particle state of the triplet operator on the right connects to the interaction and the same for the triplet on the left. In doing so, what is left from the interaction is a density operator $a^{\dag}a$ for which we will make the diagonal approximation. Of course anti-symmetrization of the two particle indices will be fully respected. We then can make a graphical representation of the interaction process contained in $\D$ as shown in Fig. \ref{diagramD}.
After the diagonalization of the matrix which implies a self-consistency on the occupancies, we can find the occupation numbers $n_h$ as
\ba
n_h=\langle a^\dagger_h a_h\rangle = \sum_{\mu} |\langle \{a^\dagger_h, q^\dagger_{h,\mu}\}\rangle|^2 =      \sum_\mu |x^\mu_h|^2
\ea
and
\ba
n_p = \langle a^\dagger_p a_p\rangle =\sum_\rho |x^\rho_p|^2
\ea
where the summation extends over all the amplitudes where $\lambda_\mu <E_F$ for hole state (or $\lambda_\rho >E_F$ for particle state). These simple expressions stem from the fact that, e.g., $a^\dagger_h $ commutes with $a^\dagger_{h}a_pa_{p'}$. 
Please notice that these occupation numbers enter also the $A$ and $B$ matrices in Eq.(\ref{normlisXY}).
Again details of the procedure will become more clear in the applications we will give below.

\noindent
Another way to find the same results for the occupancies is to define the Green Function (GF)
\ba
\G_h(\omega)=\frac{1}{\omega -\epsilon_{h} -\M_h(\omega)}
\ea
\noindent
from where we find the resonances as
\ba
\lambda_{\alpha} -\epsilon_{h} -\M_h(\lambda_{\alpha})=0 \label{eq-poles_gene}
\ea
The mass operator $\M_h$ is obtained from the eq.(\ref{odd_matrix}), eliminating the amplitude $U$,
\ba
\M_h =\sum_{pp'h',p_2p_1h_1}\C^*_{h,pp'h'}(\omega -\D)^{-1}_{pp'h',p_2p_1h_1} \C_{p_2p_1h_1,h}~
\ea
\noindent
The solution of (\ref{eq-poles_gene}) has obviously the same eigenvalues as (\ref{odd_matrix}) and then $\G_h$ can be written as 
\ba
\G_h(\omega)=\sum_{\alpha}\frac{r_{\alpha}}{\omega -\lambda_{\alpha}}\label{GeneMod}
\ea
\noindent
where 
\ba
r_{\alpha}=\frac{1}{1-\M'|_{\omega = \lambda_{\alpha}}}
\ea
\noindent
and $\M^\prime_h =d \M_h(\omega)/d\omega =-\C^\dagger(\omega -\D)^{-2} \C$.
We can easily check that $\sum_\alpha r_\alpha =1$ (the sum over all residua) and we can write the Green function dependent on time as
\ba 
i\G_h(t-t')&=&-\theta(t-t') \sum_{\alpha (\lambda_\alpha <E_F)} r_\alpha\; e^{-i\lambda_\alpha(t-t')} 
\nn
   &&           +\theta(t'-t) \sum_{\alpha (\lambda_\alpha >E_F)} r_\alpha\; e^{-i\lambda_\alpha(t-t')} 
\nn
i\G_p(t-t')&=&-\theta(t-t') \sum_{\rho (\lambda_\rho >E_F)} r_\rho\; e^{-i\lambda_\rho(t-t')} 
\nn
   &&           +\theta(t'-t) \sum_{\rho (\lambda_\rho <E_F)} r_\rho\; e^{-i\lambda_\rho(t-t')} 
\ea
Thus, we can find the s.p. occupation probabilities as
\ba 
n_h&=&-i~\underset{t'-t\rightarrow 0^{+}}{\lim }\G_h(t-t')
\nn
n_p&=&-i~\underset{t'-t\rightarrow 0^{+}}{\lim }\G_p(t-t')
\ea
Once we have the GF's, we can calculate the ground state energy in the usual way via \cite{FeterW71}
\begin{eqnarray}
E_{0} &=&-\frac{i}{2}~\underset{t'-t\rightarrow 0^{+}}{\lim }\sum\limits_{k }
\left[ i\frac{\partial }{\partial t}+\epsilon _{k}\right] \G_{k }\left( t-t'\right)   
\end{eqnarray}
\noindent
In order to test our idea, we chose two models where we know the exact solution. The first application concerns the Lipkin model as an orientation to nuclear physics. The second one focuses on solid state physics where the Hubbard model is chosen.

\section{Applications}
\subsection{The Lipkin Model}
\label{Lipkin_Model}
\begin{figure}[!]
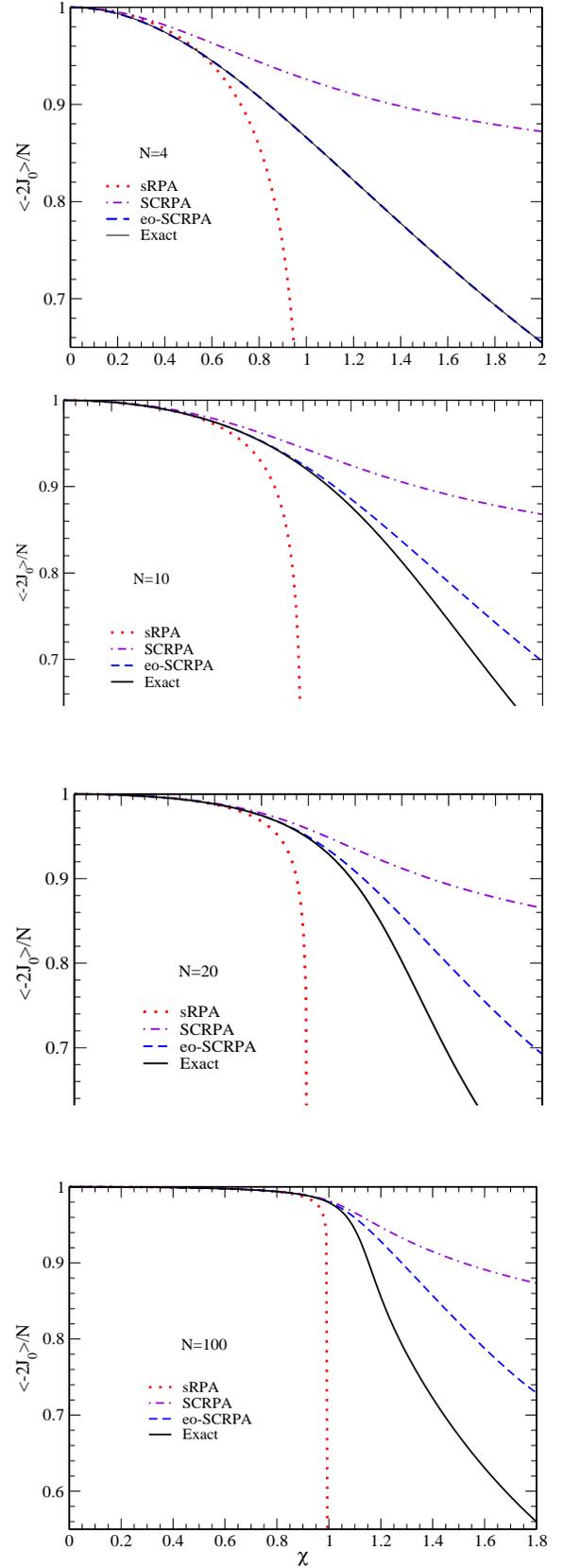

\includegraphics[width=7.5cm,height=5.45cm]{newJ0N4.eps}\vspace{0.05cm}
\includegraphics[width=7.5cm,height=5.45cm]{newJ0N10.eps}\vspace{0.05cm}
\includegraphics[width=7.5cm,height=5.45cm]{newJ0N20.eps}\vspace{0.05cm}
\includegraphics[width=7.5cm,height=5.45cm]{newJ0N100.eps}
\caption{\label{newJ0N4N10} The difference between occupation number of the two level in Lipkin model, normalized by $N$ as a function of $\chi =V(N-1)$ for $N=4,~10,~20,~100$. This with standard RPA (red dots), SCRPA (violet dashed-dot), eo-SCRPA (blue dashed line) with eom method for odd particle excitation and exact solution (full black line). Note that our approach gives the exact result for $N=4$. Also, we present the results of Catara method for $N=10$.}
\end{figure}
\begin{figure}[!]
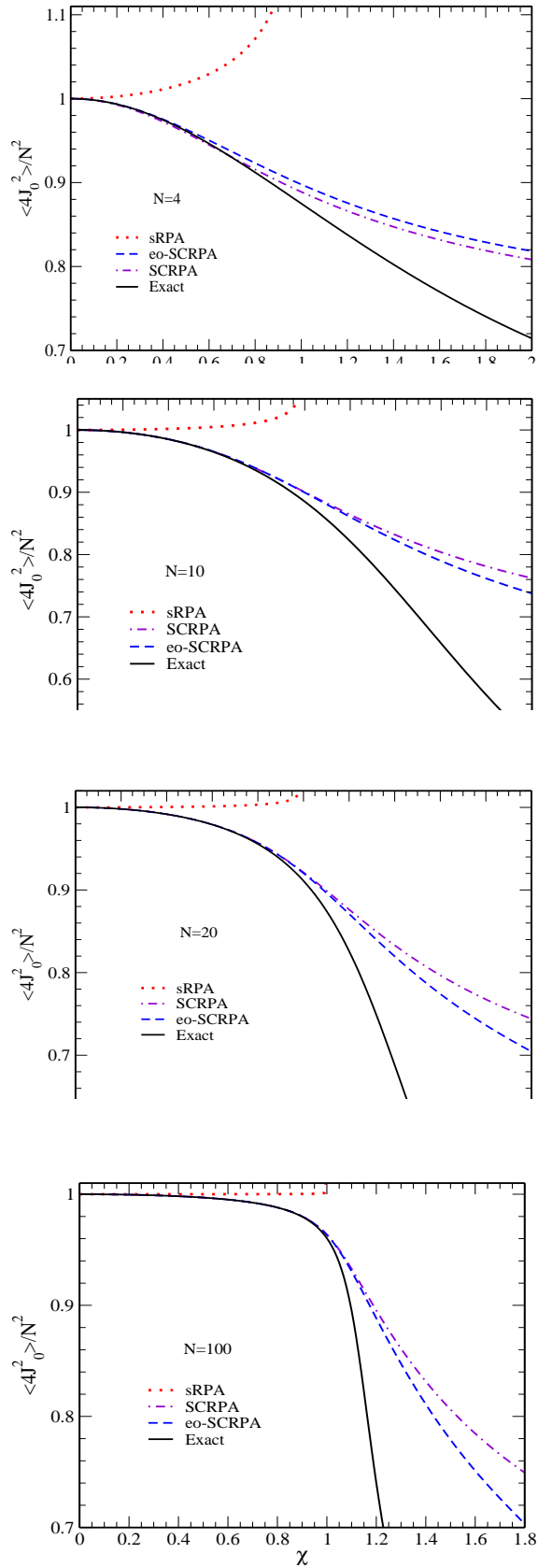

 \includegraphics[width=7.5cm,height=5.5cm]{newJ02N4.eps}\vspace{0.05cm}
 \includegraphics[width=7.5cm,height=5.5cm]{newJ02N10.eps}\vspace{0.05cm}
 \includegraphics[width=7.5cm,height=5.5cm]{newJ02N20.eps}\vspace{0.05cm}
 \includegraphics[width=7.5cm,height=5.5cm]{newJ02N100.eps}
  \caption{\label{newJ02N4N10} Same as Fig.\ref{newJ0N4N10} but for the square of the difference between occupation number of the two level in Lipkin model, normalized by $N^2$.}
\end{figure}
\begin{figure}[!]
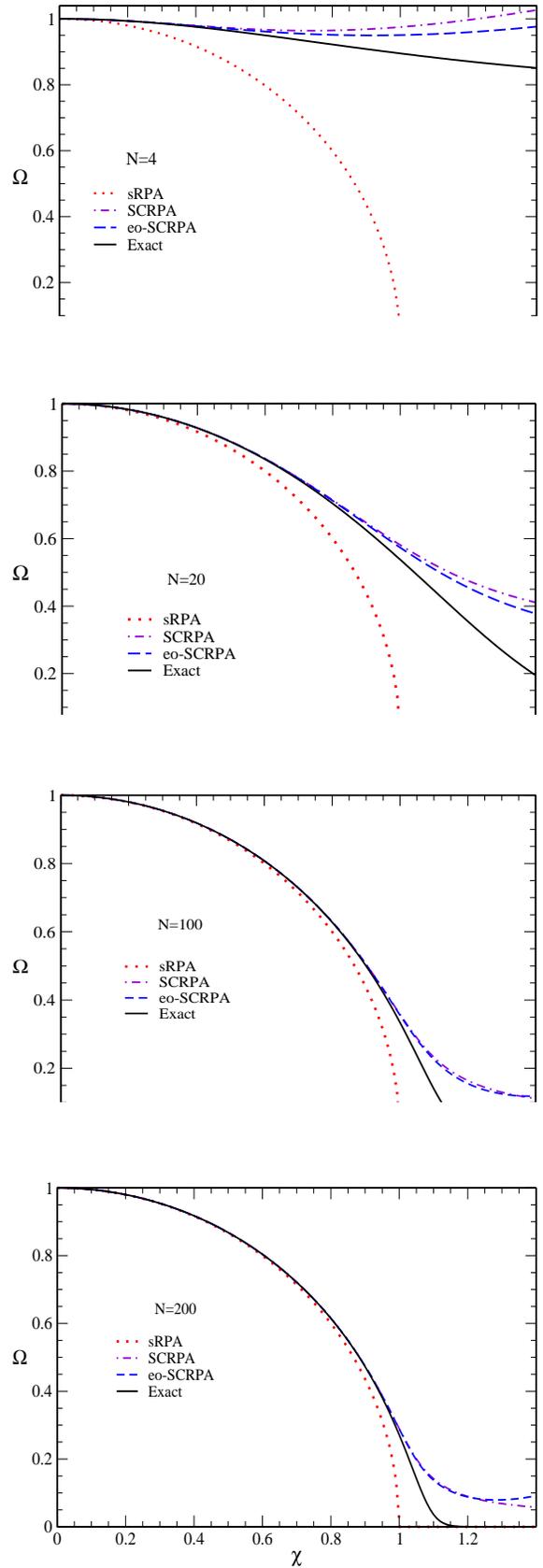

 \includegraphics[width=7.5cm,height=5.5cm]{new_AnsatzN4.eps}\vspace{0.05cm}
 \includegraphics[width=7.5cm,height=5.5cm]{new_AnsatzN20.eps}\vspace{0.05cm}
 \includegraphics[width=7.5cm,height=5.5cm]{new_AnsatzN100.eps}\vspace{0.05cm}
 \includegraphics[width=7.5cm,height=5.5cm]{new_AnsatzN200.eps}
  \caption{\label{omegaN8_density} 
Same as Fig.\ref{newJ0N4N10} but for the first excited state for $N=4,~20,~100,~200$. Please note that one may make the hypothesis that the eo-SCRPA approach becomes exact in the $N\rightarrow \infty$ limit. Also, we present the results of Catara method for $N=20$.}
\end{figure}
\begin{figure}[!]
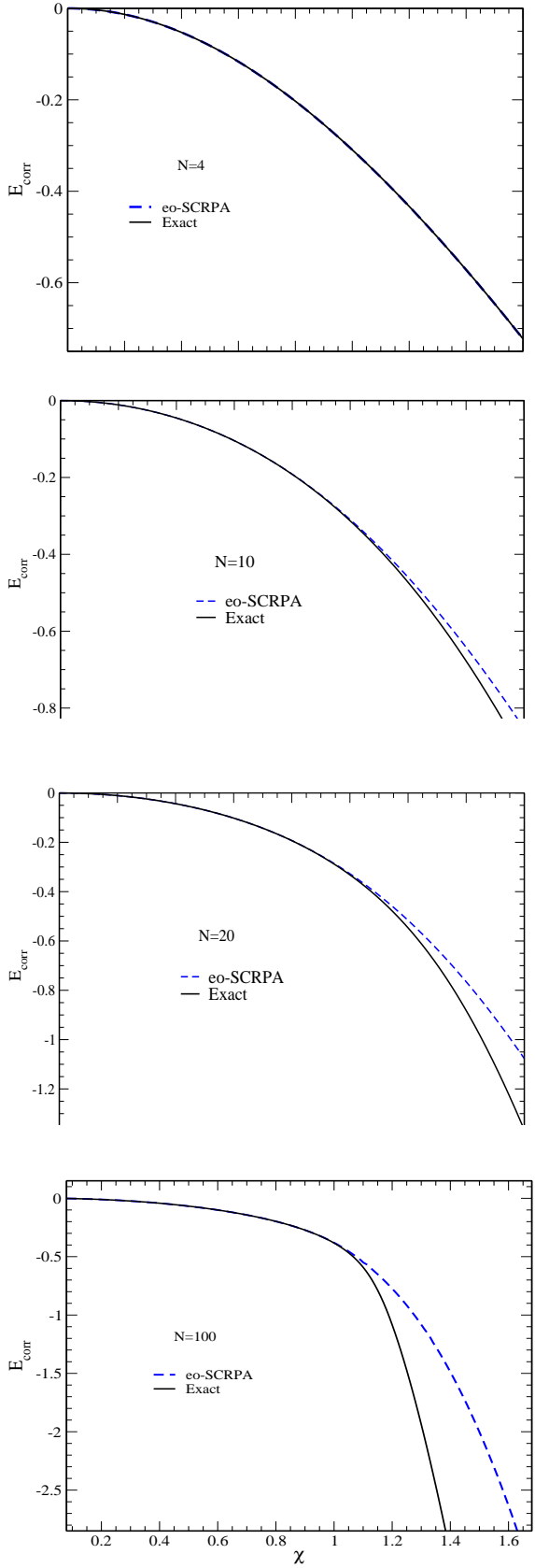

 \includegraphics[width=7.5cm,height=5.5cm]{EcorrN4GF.eps}\vspace{0.05cm}
 \includegraphics[width=7.5cm,height=5.5cm]{EcorrN10GF.eps}\vspace{0.05cm}
 \includegraphics[width=7.5cm,height=5.5cm]{EcorrN20GF.eps}\vspace{0.05cm}
 \includegraphics[width=7.5cm,height=5.5cm]{EcorrN100GF.eps}
  \caption{\label{EcorrN10GF} The correlation energy as a function of $\chi =V(N-1)$ for $N=4,~10,~20,~100$ with eo-SCRPA (blue dashed line) compared to the exact solution (full black line). Note again that for $N=4$ the exact result is obtained with our approach.}
\end{figure}
\begin{figure}[!]
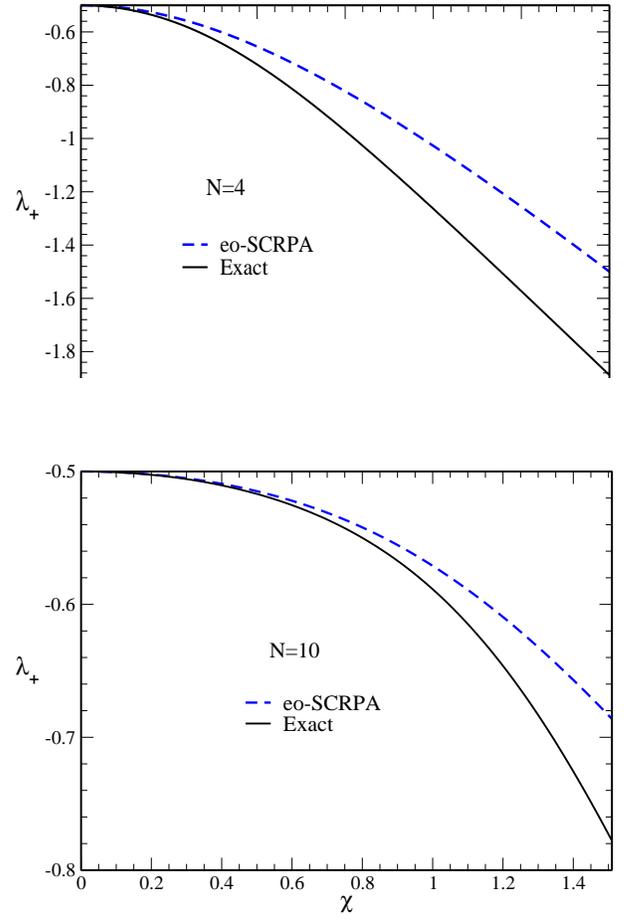

 \includegraphics[width=8cm,height=6cm]{spectN4P1oSCRPA.eps}\vspace{0.1cm}
 \includegraphics[width=8cm,height=6cm]{spectN10P1oSCRPA.eps}
  \caption{\label{SPECTN4P1oSCRPA} Excitation energy between the system $N+1$ and $N$ particles as a function of $\chi =V(N-1)$ for $N=4, ~10$ with eo-SCRPA (blue dashed line) (\ref{valpropLipk}) compared to the exact solution (full black line) $\lambda_{+}= E^{N+1}_{\alpha}-E^{N}_0$.  }
\end{figure}
\begin{figure}[ht]
 \includegraphics[width=8cm,height=6cm]{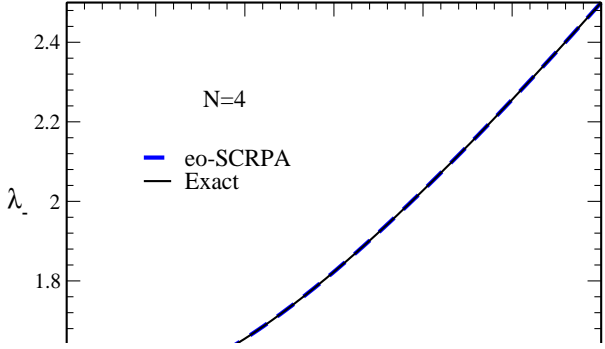}\vspace{0.1cm}
 \includegraphics[width=8cm,height=6cm]{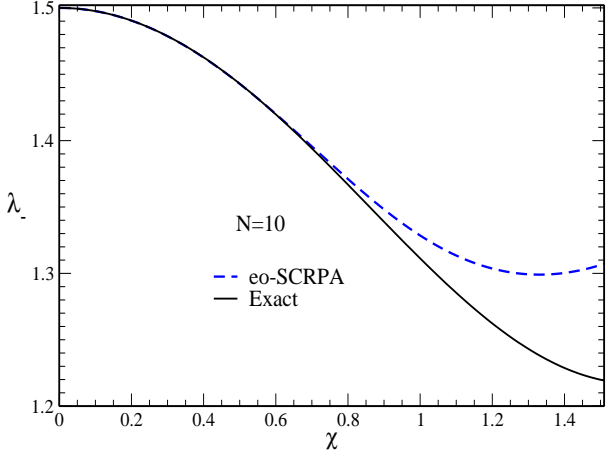}
  \caption{\label{SPECTN4M1oSCRPA} Same as Fig. \ref{SPECTN4P1oSCRPA} but for the excitation energy between the system $N-1$ and $N$ particles as a function of $\chi=V(N-1)$ for $N=4, ~10$. Note that for $N=4$ the exact result $\lambda_{-} =E^{N- 1}_{\alpha}-E^{N}_0$ is obtained with our approach eq. (\ref{valpropLipk}).  }
\end{figure}

The single-particle space of the Lipkin model consists of two fermion levels, each of which has a N-fold degeneracy see, e.g.,\cite{RingSchuck}. The upper (lower) level has the energy of $\frac{e}{2}$ ($-\frac{e}{2}$). The Hamiltonian of the Lipkin model is given by 
\begin{equation}
H=e J_0 -\frac{V}{2} \left(J^2_+ +J^2_-\right)
\label{Hlipkin}
\end{equation}
\noindent
with $e$ is the inter-shell spacing, $V$ is the coupling constant and 
\ba
J_{0}&=&\frac{1}{2}\sum_{m=1}^{N}\left(c_{1m}^{\dagger}c_{1m}-c_{0m}^{\dagger}c_{0m}\right) , 
\nn
J_{+}&=&\sum_{m=1}^{N}c_{1m}^{\dagger}c_{0m}, ~~~~~J_{-}=(\hat{J}_{+})^{\dagger },
\label{su2_operators}
\ea
\noindent
with $2J_0 = \hat{n}_1 -\hat{n}_0 $, $\hat{n}_i =\sum_m c^{\dagger}_{im} c_{im}$ and $N$ is the number of particles equivalent to the degeneracies of the shells. We consider the odd excitation operator as (\ref{odd-exc-Op}), that is
\begin{equation}
q^\dag_{\mu} =\frac{1}{N}\sum_{m} x^\mu_{0m}c_{0m} +U^\mu_{0m} J_+ c^\dagger_{1m}
\label{odd-opLipk}
\end{equation} 
with the minimisation of the sum rule in (\ref{minim-odd}). Based on the solution of the SCRPA equations \cite{jemai11,RingSchuck,jemai13} with the definition of the pair excitation operator as 
$Q^+= (XJ_+-YJ_-)/\sqrt{\langle-2J_0\rangle}$ (with $J_+=\sqrt{\langle-2J_0\rangle} (XQ^+ -YQ )$), we obtain the $X,~Y$ amplitudes as being the solutions of SCRPA equations. From the minimisation of expression (\ref{minim-odd}), we obtain a $2\times 2$ matrix eigenvalue equation (see appendix \ref{AppLipkMod}), with the Hamiltonian and norm matrices,
\begin{eqnarray}
\H_{ij}=\left(\begin{matrix}
\H_{00} & \H_{01}\\
\H_{10} & \H_{11}
\end{matrix}\right) ~~~~\mbox{and}~~~~
\N=\left(\begin{matrix}
n_{00} & n_{01}\\
n_{10} &n_{11}
\end{matrix}\right)
\end{eqnarray}
where we define the elements of the two matrices as 
\ba
n_{00}&=&\frac{1}{N}\sum_m\langle\{c_{0m},c^\dagger_{0m}\}\rangle 
\nn
n_{01}&=&n_{10}=\frac{1}{N}\sum_m\langle\{c_{0m},J_+ c^\dagger_{1m}\}\rangle 
\nn
n_{11}&=& \frac{1}{N}\sum_m\langle\{c_{1m} J_-,J_+ c^\dagger_{1m}\}\rangle
\nn
\H_{00}&=&\frac{1}{N}\sum_m\langle\{c_{0m},[H,c^\dagger_{0m}]\}\rangle 
\nn
\H_{10}&=&\H_{01}=\frac{1}{N}\sum_m\langle\{ c_{1m}J_-,[H,c^\dagger_{0m} ]\}\rangle
\nn
\H_{11}&=&\frac{1}{N}\sum_m \langle\{ c_{1m}J_-,[H,J_+c^\dagger_{1m} ]\}\rangle 
\ea 
and the corresponding secular equation 
\ba
\det\biggl\{\sum_{i'j'}\N^{-1/2}_{ii'}\H_{i'j'}\N^{-1/2}_{j'j}-\lambda I \biggr\}= 0
\label{matrixeqlipk}
\ea
\noindent
where the eigenvalues $\lambda $ are given in App. (\ref{valpropLipk}). In the above equations (\ref{matrixeqlipk}) the correlation functions are expressed by the RPA amplitudes $X, Y$ in the way it is described in section (\ref{GeneTheo}) and App. \ref{AppLipkMod}. The correlation functions which contain quadratic forms of occupation number operators as $\langle J_0J_0\rangle$ in above equation can in principle be expressed by the RPA amplitudes as well but leading to heavier expressions. Usually, we, therefore will employ the factorization approximation leading in the present case to $\langle J_0J_0\rangle \simeq \langle J_0\rangle^2$ what mostly turns out to be quite satisfactory. However, in the case of the Lipkin model one also can use the Casimir relation to close the system of equations, see App. \ref{AppLipkMod} where also  more details of the procedure are given. The results are shown in Figs. \ref{newJ0N4N10} - \ref{SPECTN4M1oSCRPA}. They concern in the order:
i) the expectation value $\langle J_0\rangle$ of the difference of populations in upper and lower level, 
ii) the square of this quantity, 
iii) the first excitation energy, 
iv) the correlation energy, and 
v) the excitation energy between the system with $N\pm1$ and $N$ particles, as $\lambda_{\pm} =E^{N\pm 1}_{\alpha}-E^{N}_0$. 
All quantities are very well reproduced throughout couplings up to the  critical value $\chi = \chi_{\rm crit.}$ where the standard RPA breaks down and the system wants to change to the 'deformed' basis. However, even values slightly beyond $\chi_{\rm crit.} =1$  are still quite acceptable. All quantities for $N=2$ are reproduced exactly. By some lucky accident the occupancies even for $N=4$ come out to be exact (as shown in Figs. \ref{newJ0N4N10}, \ref{EcorrN10GF} and \ref{SPECTN4M1oSCRPA}). 
In Fig.\ref{newJ0N4N10}, Fig.\ref{newJ02N4N10}, and Fig.\ref{omegaN8_density}, in the panels with $N=10$ and $N=20$, we also show the results of the Catara method \cite{Catara} for the calculation of the occupation numbers and first excited state
(as a reminder, let us mention that using the Catara method for the occupation numbers has been named the SCRPA method in the past; we keep the same name while getting the occupations from the selfconsistent odd RPA). 
One can thus appreciate the 
important 
improvement obtained with the method of the present work 
where even and odd RPA's are coupled.

\subsection{The Hubbard Model}
\label{HubbMod}

The Hubbard model is widely used to deal with the physics of strongly correlated electrons. Since the model can be solved exactly in one dimension (1D) and for small
cluster sizes, it is very useful for theoretical investigations \cite{jemai05}. To be precise, our "Hubbard model" is a 6-site system at half filling with periodic boundary condition, described by the usual Hamiltonian \cite{jemai05, Hubbard}: 
\begin{figure}[!]
 \includegraphics[width=5.75cm,height=3.5cm]{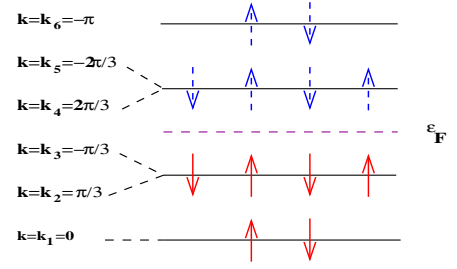}
  \caption{\label{spectrumHF6sites} Hatree Fock States at $U=0$ for the chain with 6 sites at half filling and projection of spin $m_{s} = 0$. The occupied states are represented by the full arrows and those not occupied are represented by the dashed arrows. }
\end{figure}
\begin{equation}
H=-t\sum\limits_{\langle i,j\rangle ,\sigma }c_{i\sigma }^{\dagger
}c_{j\sigma }+\frac{U}{2}\sum\limits_{i,\sigma} \hat{n}_{i,\sigma }\hat{n}_{i,-\sigma }.
\label{Hamiltonian_hubbard}
\end{equation}
Here, $\hat{n}_{i\sigma }=c_{i\sigma }^{\dagger }c_{i\sigma }$, $c_{i\sigma }^{\dagger }$ and $c_{i\sigma }$ are the creation and annihilation operators for an electron at site $i$ with spin $\sigma $, $U$
is the on-site (spin-independent) interaction, $-t$ is the hopping term of the kinetic energy. The eigenstates of the system can be expressed as linear combinations of Slater determinants.
The Hamiltonian is rewritten in plane wave basis, 
\ba
H&=&\sum_{{\bf{k}} \sigma} \varepsilon_{\bf{k}} \hat{n}_{{\bf{k}} \sigma}
+\frac U{2N}\sum_{\bf{kk'q}\sigma} a^{\dagger}_{{\bf{k}}\sigma} a_{{\bf{k+q}}\sigma}
a^{\dagger}_{{\bf{k'}} -\sigma} a_{{\bf{k'-q}} -\sigma} ~~~~~
\label{ham_imp}
\ea
with the transformation
\be
c_{j,\sigma}=\frac 1{\sqrt{N}} \sum_{\bf{k}} a_{\bf{k},\sigma} e^{-i\bf{k\,x_{j}}} ~\mbox{,} \label{6sit_trans_conssym0ca}
\ee
where $\hat{n}_{\bf{k}, \sigma} = a^{\dagger}_{\bf{k}, \sigma} a_{\bf{k}, \sigma}$, $\varepsilon _{\bf{k}}= -2t \cos\left(ka\right)$, which are, respectively, the number operator of particles of the mode $({\bf{k}},\,\sigma)$ and the energies of one particle on a lattice with a the parameter of the lattice which is taken as $a=1$. For a problem with $N$sites, the condition of periodicity is given by $c_{N+1,\sigma}=c_{1,\sigma}$. This implies that $e^{-ik\,N}=1$, hence the values taken by $k$ will be $k=\frac {2\,\pi}{N}\,n$. In addition, the first Brillouin zone is defined on the field where $-\pi\leqslant k <\pi$, which gives us the values of $n$ as $\frac{-N}{2}\leqslant n <\frac{N}{2} $.

For the six sites, we have the possible states with the following wave vectors:
\ba
k_1=0,~k_3=-k_2=\frac{\pi}{3} ,~k_5=-k_4=\frac{2\pi}{3} ,~k_6=-\pi
\ea 
and with the kinetic energies (see Fig.\ref{spectrumHF6sites}), respectively,
\ba
\varepsilon_{k_{6}}=-\varepsilon_{k_{1}}=2\,t,~~
\varepsilon_{k_{4}}=\varepsilon_{k_{5}}=-\varepsilon_{k_{2}}=-\varepsilon_{k_{3}}=t .
\ea 
\noindent
The transfer wave vector($q_{ph}=k_p-k_h$) takes the possible values as shown in the Table \ref{table1}.
\begin{table}[h]
\begin{tabular}{|c|c|c|}
\hline
$q =\pm\frac{2\pi}{3}$  & $q =\pm \pi$ & $q =\pm \frac{\pi}{3}$
\\ \hline
$51\rightarrow q_{51} =+\frac{2\pi}{3}$ & $61\rightarrow q_{61}  =-\pi$ & $42\rightarrow q_{42} =-\frac{\pi}{3}$
\\ \hline
$63\rightarrow q_{63} =+\frac{2\pi}{3}$ & $52\rightarrow q_{52}  =+\pi$ & $53\rightarrow q_{53} =+\frac{\pi}{3}$
\\ \hline
$41\rightarrow q_{41} =-\frac{2\pi}{3}$ & $43\rightarrow q_{43}  = -\pi$ &
\\ \hline
$62\rightarrow q_{62} =-\frac{2\pi}{3}$ &  &
\\ \hline
\end{tabular}
\caption{\label{table1} The various momentum transfers in the 6 sites case.}
\end{table}
\begin{figure}[!]
 \includegraphics[width=8cm,height=6.5cm]{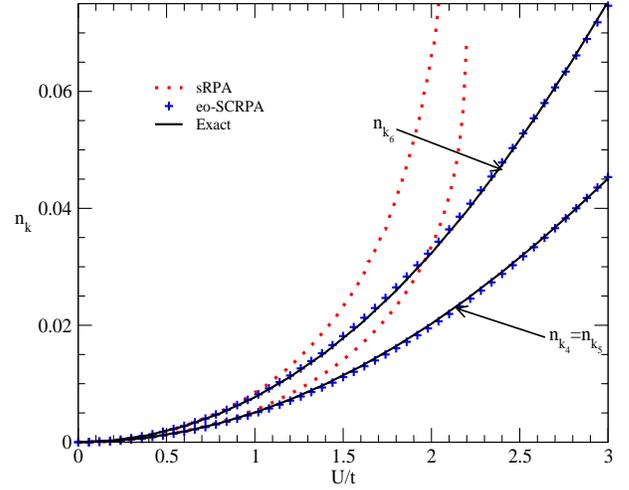}
  \caption{\label{Hcomplet} Occupation numbers as function of the interaction $U/t$ for various values of the momenta $k_6=-\pi$, $k_5=-2\pi/3$, $k_4=2\pi/3$ for states above the Fermi level. Notice that the modes $k_4 = 2\pi/3$ and $k_5 =-2\pi/3$ are degenerate. For each approximation, sRPA (red dots) and eo-SCRPA (blue crosses), are  compared to the exact solution (full black line).  Also we have $n_{k_1}=1-n_{k_6}$ and $n_{k_2}=n_{k_3}=1-n_{k_4}=1-n_{k_5}$ }
\end{figure}
\begin{figure}[!]
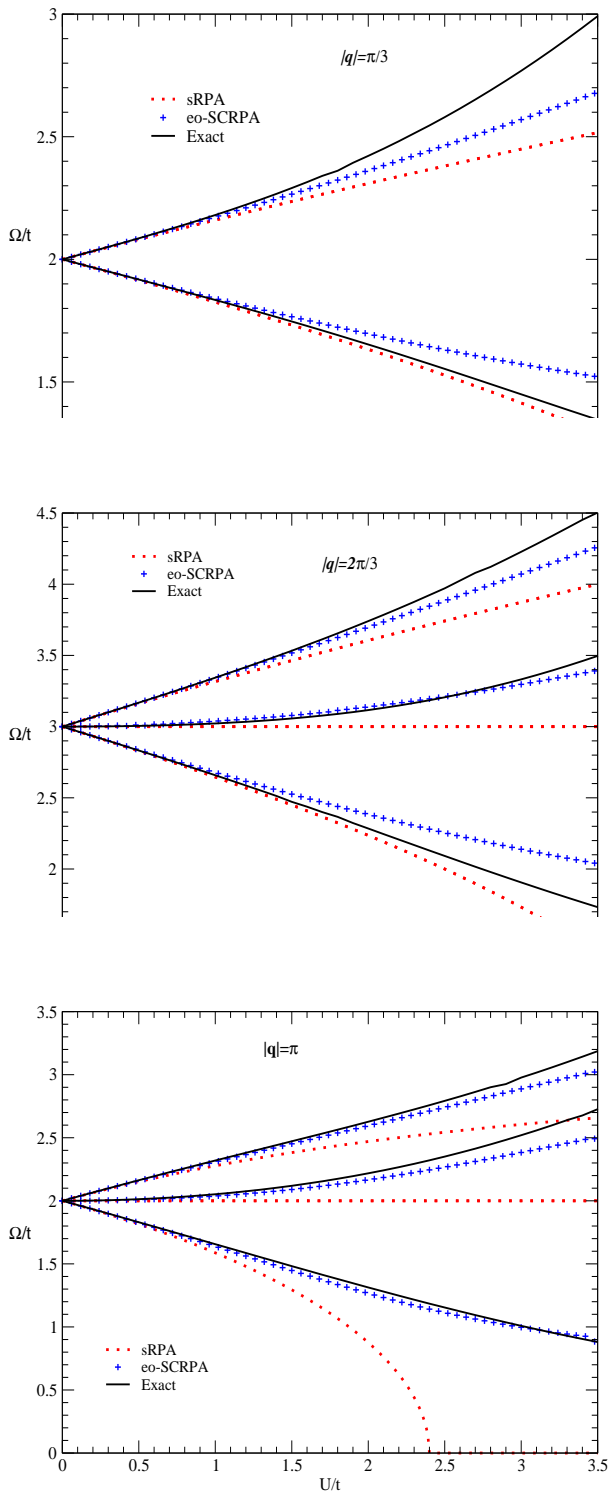

 \includegraphics[width=8cm,height=6.5cm]{Rep1p3.eps}\vspace{0.1cm}
 \includegraphics[width=8cm,height=6.5cm]{Rep2p3.eps}\vspace{0.1cm}
 \includegraphics[width=8cm,height=6.5cm]{Rep1p1.eps}
  \caption{\label{rep1p1} Same as Fig.\ref{Hcomplet} but for the energies of excited states for different channels $|q| =\pi/3$, $2\pi/3$ and $\pi$. }
\end{figure}
\begin{figure}[!]
 \includegraphics[width=8cm,height=6.5cm]{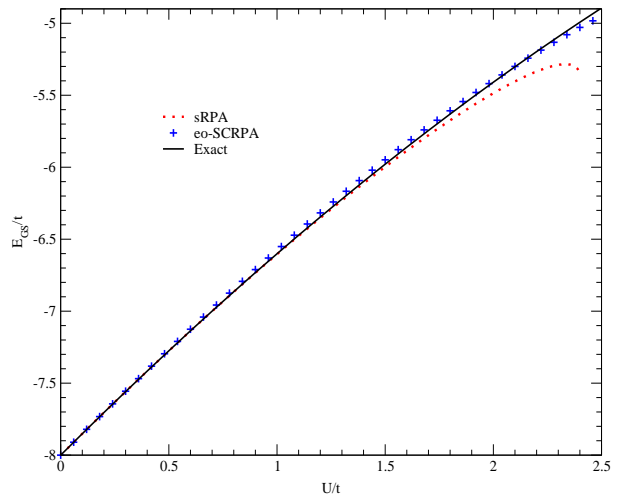}
  \caption{\label{EGS_Hubb6ph} Same as Fig.\ref{Hcomplet} but for the ground state energy. }
\end{figure}

At this point we proceed exactly as in the case of the Lipkin model: The excitation operator for the even system is given by
\ba
Q^{\dagger}_{\nu} = \sum_{ph\sigma}\X^{\nu}_{ph\sigma}\, K^{+}_{ph\sigma}-\Y^{\nu}_{ph\sigma}\, K^{-}_{hp\sigma}
\label{op-dexqp6s4}
\ea
with $~ K^{\pm}_{ph\sigma }=J^{\pm}_{ph\sigma}/\sqrt{N_{ph\sigma}} $, $~J^{+}_{ph\sigma}=a^{\dag}_{p\sigma}a_{h\sigma}$, $~N_{ph\sigma}=n_{h\sigma}- n_{p\sigma}$. With the inversion
\ba
J^{-}_{hp\sigma} & = & \sqrt{N_{ph\sigma}}\;\sum\limits_{\nu}\;\left(\;\X^{\nu}_{ph\sigma} \; Q_{\nu} + \Y^{\nu}_{ph\sigma} \; Q_{\nu}^{\dagger}\;\right)
\nonumber \\
J^{+}_{ph\sigma} & = & \sqrt{N_{ph\sigma}}\sum\limits_{\nu}\;\left(\;\Y^{\nu}_{ph\sigma} \; Q_{\nu} + \X^{\nu}_{ph\sigma} \; Q_{\nu}^{\dagger}\;\right)~.
\label{inversion_1}
\ea
we can calculate the mean values needed for the matrix elements of the SCRPA equations for the even particle number case
\ba
\langle J^{+}_{p'h'\sigma '}\,J^{-}_{hp\sigma}\rangle & = &
\sqrt{N_{p'h'\sigma '} N_{ph\sigma}}\; \sum\limits_{\nu}\;\Y^{\nu}_{p'h'\sigma '}\; \Y^{\nu}_{ph\sigma}
\mbox{,}
\nonumber \\
\langle J^{-}_{h'p'\sigma '}\,J^{+}_{ph\sigma}\rangle & = &
\sqrt{N_{p'h'\sigma '} N_{ph\sigma}}\; \sum\limits_{\nu}\;\X^{\nu}_{p'h'\sigma '}\; \X^{\nu}_{ph\sigma}
\mbox{,}
\nonumber \\
\langle J^{+}_{p'h'\sigma '}\,J^{+}_{ph\sigma}\rangle & = &
\sqrt{N_{p'h'\sigma '} N_{ph\sigma}}\; \sum\limits_{\nu}\;\Y^{\nu}_{p'h'\sigma '}\; \X^{\nu}_{ph\sigma}
\mbox{,}
\nonumber \\
\langle J^{-}_{h'p'\sigma '}\,J^{-}_{hp\sigma}\rangle & = &
\sqrt{N_{p'h'\sigma '} N_{ph\sigma}}\; \sum\limits_{\nu}\;\X^{\nu}_{p'h'\sigma '}\; \Y^{\nu}_{ph\sigma}
\mbox{,}~~~\label{foctcorr_jj}
\ea
where we replaced the "ph" operators by the RPA creation and destruction operators from the inversion (\ref{invQ}) and then commute the $Q$ operators to the right until they kill the ground state. All matrices become functional of the occupancies $n_h$ and $n_p$ and $X, Y$ amplitudes in analogy to what was the case in the Lipkin model and, thus, the diagonalisation process implies at the same time an iteration on the occupancies and the amplitudes.\\ 
For the odd particle number case, we make again the following ansatz
\begin{eqnarray}
q^\dag_{h,\mu} &=& x^\mu_{h} a_{h+} +\sum_{p'ph} U^\mu_{p'ph} a^\dagger_{p'+} J^+_{ph-}
\nn
q^\dag_{p,\rho}&=& x^\rho_{p} a^\dagger_{p+} + \sum_{p'h'h} U^\rho_{p'h'h} a^\dagger_{h+}J^-_{h'p'-}  ~.
\end{eqnarray}
From there, we can, as outlined in the general section II and as just now for the case of the Lipkin model calculate the occupation numbers. For more details, see App. \ref{AppHubbMod}. The results for the occupation numbers are again very satisfying, see Fig.~\ref{Hcomplet}. Also the excitation energies of the even particle number system, see Fig.~\ref{rep1p1} are very well reproduced. In Fig.~\ref{EGS_Hubb6ph} we show the ground state energies for the exact case compared to the eo-SCRPA solution.\\
One should notice that there is barely an improvement using the eo-SCRPA versus the standard SCRPA because the latter produced already excellent results. So, we do not show the old SCRPA results again. It is not quite clear why there is this difference between the Lipkin and Hubbard models. Probably the fact that in Lipkin model, contrary to the Hubbard model, one uses {\it collective} ph operators makes it more difficult to fullfill the Pauli principle. So the performance of one or the other approach seems to depend on the
situation.

\section{Conclusion}
In this work, we coupled even and odd particle number RPA self-consistently. Both systems are based on the same correlated RPA ground state. From the odd system, we get the occupation numbers, odd particle excitation energies, and the ground state energies whereas from the even SCRPA equations we get the excitation energies of the even system and transition probabilities. 
To make things clear, we should mention again that the SCRPA employed here has the same mathematical structure as the one used before \cite{jemai05}, only the single particle occupation probabilities are now calculated via the odd selfconsistent RPA whereas they were obtained before via the so-called Catara method \cite{Catara}.
Both even and odd systems are coupled through non-linear equations 
which both contain the RPA amplitudes $X, Y$ and the s.p. occupation numbers $n_k$ in a non-linear way.
We called this system of equations 'even-odd' SCRPA (eo-SCRPA). Applications to the Lipkin model and a six sites Hubbard ring at half filling gave very satisfying results for all quantities. The equations are relatively complex due to their non-linearity but they should be solvable with modern computers for realistic problems 
such as, e.g., the calculation of collective states in nuclei. The equations to be solved seem not to be of higher numerical complexity than, e.g., the Brueckner Hartree-Fock equations which have been solved a number of times for nuclei.
The coupling of even and odd RPA's has a couple of advantages: it gives richer results, i.e., excitation energies of even and odd particle number systems; there is a natural way how to obtain the ground state energy via the s.p. Green's function and, last but not least, the results seem to be promising.

\section{Acknowledgements}

We are grateful for long-standing collaboration on SCRPA with D. Delion, J. Dukelsky, and M. Tohyama.

\appendix

\section{ Equation of Motion for odd particle number operator for Lipkin Model}
\label{AppLipkMod}
\noindent
We consider the odd excitation operator as
\begin{equation}
q^\dag_{\mu} =\frac{1}{N}\sum_{m} x^\mu_{0m}c_{0m} +U^\mu_{0m} J_+ c^\dagger_{1m}
\label{odd-ansatz22}
\end{equation}
\noindent
and the coefficients will be determined from minimisation of expression (\ref{minim-odd}).
Based on the solution of the SCRPA equations with the definition of the pair excitation operator as 
$Q^\dag= (XJ_+-YJ_-)/\sqrt{\langle-2J_0\rangle}$ (with $J_+=\sqrt{\langle-2J_0\rangle} (XQ^\dag -YQ )$), the $X,~Y$ amplitudes are the solutions of the  SCRPA equations  with $H$ of the Lipkin Hamiltonian (\ref{Hlipkin}). From the minimisation of (\ref{minim-odd}), we obtain a $2\times 2$ matrix eigenvalue equation. 
\noindent
The norm matrix is given by
\begin{eqnarray}
n_{00}&=&\frac{1}{N}\sum_m\langle\{c_{0m},c^\dagger_{0m}\}\rangle =1
\nn
n_{01}&=& n_{10} = \frac{1}{N}\sum_m\langle\{c_{0m},J_+ c^\dagger_{1m}\}\rangle =0
\nn
n_{11}&=& \frac{1}{N}\sum_m\langle\{c_{1m} J_-,J_+ c^\dagger_{1m}\}\rangle 
\nn
&=&-\frac{1}{N}\left(N-2\right)\left(1+2Y^2 \right)\langle J_0\rangle 
+\frac{2}{N} \langle J_0J_0\rangle  ~~~~
\end{eqnarray}
where we have used the inversion (\ref{invQ}) and the killing condition $Q|0\rangle = 0$.
\noindent
For the first Hamiltonian element we have
\begin{eqnarray}
\H_{00} =\frac{1}{N}\sum_m \langle\{c_{0m},\left[H,c^\dagger_{0m}\right]\}\rangle = -\frac{e}{2} 
\end{eqnarray}
\noindent
and for the off diagonal elements  
\begin{eqnarray}
\H_{10} &=& \H_{01}= \frac{1}{N}\sum_m \langle\{ c_{1m}J_-,[H,c^\dagger_{0m} ]\}\rangle
\nn
&=&-\frac{e}{2N}\sum_m\langle\{ c_{1m}J_-,c^\dagger_{0m}\}\rangle 
\nn
&&~~~-\frac{V}{N}\sum_m\langle\{ c_{1m} J_-, J_+ c^\dagger_{1m}\}\rangle
\nn
&=&-Vn_{11}
\end{eqnarray}
\noindent
And the anti-commutator for $\H_{11}$ is given by
\begin{eqnarray}
\H_{11}&=&\frac{1}{N}\sum_m \langle\{ c_{1m}J_-,[H,J_+c^\dagger_{1m} ]\}\rangle 
\\
&=&\frac{3e}{2} n_{11} +V(2-\frac{8}{N})[\langle J_-J_-\rangle
 +\langle J_0J_- J_-\rangle ]
\nn
&=&\frac{3e}{2} n_{11} -2VXY(2 -\frac{8}{N})[(1 +2Y^2)\langle J_0\rangle
+ \langle J^2_0\rangle]\nonumber
\end{eqnarray}
with, for example
\ba
\langle J_0 J_- J_-\rangle =-4XY^3 \langle J_0\rangle -2XY \langle J_0J_0\rangle
\ea
where we have again used the inversion (\ref{invQ}) and the killing condition $Q|0\rangle = 0$.

The correlation functions which contain quadratic forms of occupation number operators as $\langle J_0J_0\rangle$ in above equation can in principle be expressed by the RPA amplitudes as well but leading to heavier expressions. Usually, we, therefore will employ the factorization approximation leading in the present case to $\langle J_0J_0\rangle \simeq \langle J_0\rangle^2$. However, in the Lipkin model one also can use the Casimir relation 
\ba
\langle 4 J_0J_0\rangle &=&N(N+2) +4\langle J_0\rangle - 4\langle J_+J_-\rangle
\label{eqJ0_exp3}
\ea
Then all matrix elements $\H_{ij}$ become functions of $\langle J_0\rangle$ and the RPA amplitudes $X, Y$. The eigenvalue problem can therefore be solved leading to a self-consistency problem for $\langle J_0\rangle$ and the RPA amplitudes which are obtained from the SCRPA equations (\ref{elemt-mat}) \cite{jemai13}. The occupation numbers are then given by 
\begin{eqnarray}
n_{0} &=& N \frac{\lambda _- -\H_{11}/n_{11}}{\lambda_{-} -\lambda _+}
~~~\mbox{and}~~~
n_1 = N-n_0~~~
\end{eqnarray}
where $\lambda_{\pm}$ are the eigenvalues of the $2\times 2$ matrix problem,
\begin{eqnarray}
\lambda _\pm &=& -\frac{e}{2} +\beta\pm \sqrt{\beta ^2+V^2 n_{11}}
\label{valpropLipk}
\end{eqnarray}
\noindent
with $\beta =e- VXY(N-4)-VXY(N-4)(1+2Y^2)\frac{\langle J_0\rangle}{n_{11}}$. Thus,
\begin{eqnarray}
\langle -2 J_0\rangle = n_0 - n_1 = 2 n_0 -N
\label{2j0Wx0}
\end{eqnarray}

\section{Equation of Motion for Hubbard Model}
\label{AppHubbMod}
\noindent
For the Hubbard model (\ref{ham_imp}) we define the odd excitation operator as in Eq.(\ref{odd-exc-Op}),
\begin{eqnarray}
q^\dag_{h,\mu} &=& x^\mu_{h} a_{h+} +\sum_{p'ph} U^\mu_{p'ph} a^\dagger_{p'+} J^+_{ph-}
\nn
\nn
q^\dag_{p,\rho}&=& x^\rho_{p} a^\dagger_{p+} + \sum_{p'h'h} U^\rho_{p'h'h} a^\dagger_{h+}J^-_{h'p'-}  ~.
\end{eqnarray}
\noindent 
with $J^+_{ph-} =a^\dagger_{p-} a_{h-}$ and $\sigma =\uparrow, \downarrow=+,-$.
Remembering the  notations for the occupation probabilities
\begin{eqnarray}
n_{k\sigma}=\langle\hat{n}_{k\sigma}\rangle &=& \langle a^\dagger_{k\sigma} a_{k\sigma}\rangle , 
\end{eqnarray}
we have $n_{k_2\sigma} =n_{k_3\sigma}$, $n_{k_4\sigma} =n_{k_5\sigma}$, $n_{k_2\sigma} =1-n_{k_3\sigma}$ and $n_{k_1\sigma} =1-n_{k_6\sigma}$. This gives 
\begin{eqnarray}
\H_{11}=\langle\{a_{k_1+},[H, a^\dagger_{k_1+}]\}\rangle = \epsilon_{k_1} =-2t +U/2
\end{eqnarray}
\noindent
The term without interaction $H_{0}=\sum_{k\sigma} \varepsilon_{k} \hat{n}_{k\sigma} $ is given by 
\begin{eqnarray}
\langle \{ a_{p'+} J^-_{hp-},[H_{0}, a^\dagger_{p'+} J^+_{ph-}] \} \rangle 
&=&(\varepsilon_{p} -\varepsilon_{h} +\varepsilon_{p'})\N_{p'ph} \nn
\ea
with $\N_{p'ph}=\langle (1-\hat{n}_{p'+})(-2 J^0_{ph,-})\rangle +\langle J^+_{ph,-} J^{-}_{hp,-} \rangle$. The term in the Hamiltonian for the transfer $q=0$, $H_{q=0}=\frac U{6}\sum_{kk'} \hat{n}_{k+} \hat{n}_{k' -} $ leads to
\ba
\langle \{ a_{p'+} J^-_{hp-},[H_{q=0}, a^\dagger_{p'+} J^+_{ph-}] \} \rangle  &=& \frac{U}{2}\N_{p'ph} 
\end{eqnarray}
with $\sum_k \hat{n}_{k\sigma} =\sum_p \hat{n}_{p\sigma} +\sum_h \hat{n}_{h\sigma} =3 $ in the half-filled case. Now let us calculate the elements $\C_{p'ph}$ for the first row (or column) as
\begin{widetext}
\ba
\sqrt{\N_{p'ph}}\C^*_{p'ph,h_1} &=&\langle\left\{a_{p'+} J^-_{hp-}, \left[H,a^\dagger_{h_1+} \right]\right\}\rangle
\nn
&=&\frac{U}{6}\biggr\{\langle a^\dagger_{h_1-q+}a_{p'+} a^\dagger_{h+q-}a_{p-}\rangle
-\langle a^\dagger_{h_1-q+}a_{p'+} a^\dagger_{h-}a_{p-q-}\rangle
+\sum_{k} \langle J^-_{hp-} a^\dagger_{k-} a_{k+p'-h_1-}\rangle
 \biggl\}
\ea
The elements of the matrix except the first row (or column) are given as follows 
\ba
&&\sqrt{\N_{p'ph}\N_{p''p_1h_1}}\D_{p'ph,p''p_1h_1} = \langle \{ a_{p''+} J^-_{h_1p_1-},[H, a^\dagger_{p'+} J^+_{ph-}] \} \rangle
\nn
&=&(\epsilon_{p}+\epsilon_{p'}-\epsilon_{h})\dl_{p'p''}\biggl\{\langle J^-_{h_1p_1-}  J^+_{ph-}\rangle  +\dl_{hh_1}\dl_{pp_1}\langle \hat{n}_{p'+}(\hat{n}_{p-}-\hat{n}_{h-})\rangle \biggr\}
\nn
&&+\frac{U}{6}\dl_{p'p''}\biggl\{\sum_{kq}
 \langle a^\dagger_{k+}a_{k+q+} J^-_{h_1p_1-} (a^\dagger_{p+q-} a_{h-}-a^\dagger_{p-} a_{h-q-})\rangle \biggr\}
\nn
&&+\frac{U}{6}\dl_{p'p''}\dl_{hh_1}\biggl\{\sum_{kq}
 \langle a^\dagger_{k+}a_{k+q+} \hat{n}_{p'+} a^\dagger_{p+q-}a_{p_1-}\rangle 
-\sum_{k}\langle a^\dagger_{k+}a_{k+p_1-p+} \hat{n}_{p'+} \hat{n}_{h-}\rangle \biggr\}
\nn
&&+\frac{U}{6}\dl_{p'p''} \dl_{pp_1}\biggl\{\sum_{kq}
\langle a^\dagger_{k+}a_{k+q+} \hat{n}_{p'+} a^\dagger_{h_1-}a_{h-q-}\rangle 
-\sum_{k}\langle a^\dagger_{k+}a_{k+h-h_1+} \hat{n}_{p'+} \hat{n}_{p-}\rangle \biggr\}
\nn
&&+\frac{U}{6}\dl_{pp_1}\dl_{hh_1}\biggl\{\sum_{kq}
\langle a^\dagger_{p'-q+}a_{p''+}\hat{n}_{p-}a^\dagger_{k-}a_{k-q-}\rangle
-\sum_{kq}\langle a^\dagger_{p'-q+}a_{p''+}\hat{n}_{h-}a^\dagger_{k-}a_{k-q-}\rangle 
\biggr\}
\nn
&&+\frac{U}{6}\biggl\{\sum_{q}
\langle a_{p''+q+}a^\dagger_{p'+} J^-_{h_1p_1-} (a^\dagger_{p+q-} a_{h-}- a^\dagger_{p-} a_{h-q-})\rangle
\nn
&&~~~~~+\sum_{q}\langle a^\dagger_{p'-q+}a_{p''+} J^+_{ph-} (a^\dagger_{h_1+q-}a_{p_1-}-a^\dagger_{h_1-}a_{p_1-q-})\rangle 
+\sum_{k}\langle J^-_{h_1p_1-} J^+_{ph-} a^\dagger_{k-} a_{k-p'+p''-}\rangle\biggr\}
\label{triplet}
\ea
\end{widetext}
\noindent
In the following, as already discussed several times, we retain from (\ref{triplet}) only those terms where the particle states of the left and right triple operators in $\D$ connect to the interaction. The remaining density operator from the interaction is approximated by its diagonal form. This leads to expressions evaluated in (\ref{ExampTriplet}) below. First let us discuss what kind of terms we are neglecting in this way.
It should be noted that the terms of type $\langle J^{\pm}_{ph}J^{\pm}_{p'h'}J^{\pm}_{p''h''}\rangle =0$, $\langle J^{\pm}_{ph}S_{p_1p_2}J^{\pm}_{p''h''}\rangle$ are probably small (with $S_{p_1p_2}=a^\dagger_{p_1} a_{p_2}$ for $p_1\neq p_2$) and $\langle J^{\pm}_{ph}S_{h_1h_2}J^{\pm}_{p''h''}\rangle$ also small (with $S_{h_1h_2}=a^\dagger_{h_1} a_{h_2}$ for $h_1\neq h_2$)  in eq.(\ref{triplet}). As shown in \cite{jemai13}, the term $\langle SJ\rangle =0$ and $\langle SS\rangle$  are small. Only the terms non-zero in eq.(\ref{triplet}) like $\langle J^{\pm}_{ph+}n_{k\pm} J^{\pm}_{p'h'-}\rangle $ which can be calculated as shown in (\ref{ExampTriplet}) are kept. With the short hand notation $ph\sigma \equiv i$, $k\sigma \equiv k$, $\hat{N}_i=\hat{n}_{h\sigma} -\hat{n}_{p\sigma}$ and $N_i=n_{h\sigma} -n_{p\sigma}$, we can evaluate the following terms
\begin{widetext}
\ba
\langle J^-_{i}\hat{n}_{k} J^-_{j}\rangle &=&\sqrt{N_iN_j} \sum_{\nu,\nu '} X^\nu_i Y^{\nu '}_j \langle Q_\nu \hat{n}_{k} Q^\dag_{\nu '} \rangle
\nn
&=&\sqrt{N_iN_j} \sum_{\nu,\nu '} X^\nu_i Y^{\nu '}_j \left( X^\nu_i X^{\nu '}_j  -Y^\nu_i Y^{\nu '}_j \right) 
+ \sum_{\nu} X^\nu_i Y^{\nu }_j \sum_{l} \left( |X^\nu_l|^2 -|Y^\nu_l|^2\right) \langle \hat{n}_{k} \hat{N}_l \rangle
\nn
\langle J^+_{i}\hat{n}_{k} J^-_{j}\rangle &=&\sqrt{N_iN_j} \sum_{\nu,\nu '} Y^\nu_i Y^{\nu '}_j \langle Q_\nu \hat{n}_{k} Q^\dag_{\nu '} \rangle
\nn
&=&\sqrt{N_iN_j} \sum_{\nu,\nu '} Y^\nu_i Y^{\nu '}_j \left( X^\nu_i X^{\nu '}_j  -Y^\nu_i Y^{\nu '}_j \right) 
+ \sum_{\nu} Y^\nu_i Y^{\nu }_j \sum_l \left( |X^\nu_l|^2 -|Y^\nu_l|^2\right) \langle \hat{n}_{k} \hat{N}_{l} \rangle ~~~~~
\nn
\langle J^-_{i}\hat{n}_{k} J^+_{j}\rangle &=&\sqrt{N_iN_j} \sum_{\nu,\nu '} X^\nu_i X^{\nu '}_j \langle Q_\nu \hat{n}_{k} Q^\dag_{\nu '} \rangle
\nn
&=&\sqrt{N_iN_j} \sum_{\nu,\nu '} X^\nu_i X^{\nu '}_j \left( X^\nu_i X^{\nu '}_j  -Y^\nu_i Y^{\nu '}_j \right) 
+ \sum_{\nu} X^\nu_i X^{\nu}_j \sum_l \left( |X^\nu_l|^2 -|Y^\nu_l|^2\right) \langle \hat{n}_{k} \hat{N}_{l} \rangle ~~~~~
\label{ExampTriplet}
\ea
where we used $Q^\dag_\nu =\sum_l (X^\nu_l J^+_l +Y^\nu_l J^-_l)N^{-1/2}_l $ and the commutators
\ba
\left[Q_\nu ,\hat{n}_{k}\right]&=&+N^{-1/2}_l \left( X^\nu_l J^-_l -Y^\nu_l J^+_l\right)
\nn
\left[Q_\nu ,\hat{n}_{k}\right]&=&-N^{-1/2}_l \left( X^\nu_l J^-_l -Y^\nu_l J^+_l\right)
\nn
\left[Q_\nu ,Q^\dag_{\nu '}\right]&=&\sum_l\left( X^\nu_l X^{\nu '}_l -Y^\nu_l Y^{\nu '}_l\right)\hat{N}_{l} N^{-1}_l 
\nn
&=&\delta_{\nu,\nu '} \sum_l\left(|X^\nu_l|^2 -|Y^\nu_l|^2\right)\hat{N}_{l} N^{-1}_l.
\ea
This entails, $\langle [Q_\nu ,Q^\dag_{\nu '}]\rangle =1$ and 
\ba
\langle Q_\nu \hat{n}_{k} Q^\dag_{\nu '} \rangle &=&
N^{-1/2}_i \left( X^\nu_i\langle J^-_i Q^\dag_{\nu '} \rangle -Y^\nu_i\langle J^+_i Q^\dag_{\nu '} \rangle \right) 
+\langle \hat{n}_{k} Q_\nu  Q^\dag_{\nu '} \rangle
\nn 
&=&\left( X^\nu_i X^{\nu '}_i  -Y^\nu_i Y^{\nu '}_i \right) 
+\delta_{\nu,\nu '}  \sum_l N_l^{-1}\left( X^\nu_l X^{\nu}_l-Y^\nu_l Y^{\nu}_l\right)\langle \hat{n}_{k} \hat{N}_{l} \rangle
\ea
\end{widetext}

\bibliographystyle{refer}
\bibliography{articles}

\end{document}